\documentclass[12pt,preprint]{aastex}

\begin{document}

\title{COMPONENT MASSES OF THE YOUNG SPECTROSCOPIC BINARY UZ TAU E}

\author{{L. PRATO\altaffilmark{1}, M. SIMON\altaffilmark{2},
T. MAZEH\altaffilmark{3},
S. ZUCKER\altaffilmark{3}, AND I. S. MCLEAN\altaffilmark{1}}}

\altaffiltext{1}{Department of Physics and Astronomy, UCLA,
Los Angeles, CA 90095-1562; lprato@astro.ucla.edu}
\altaffiltext{2}{Department of Physics and Astronomy, SUNY,
Stony Brook, NY 11794-3800}
\altaffiltext{3}{Department of Physics and Astronomy, Tel Aviv University,
Tel Aviv, Israel}

\begin{abstract}

We report estimates of the masses of the component stars
in the pre$-$main-sequence spectroscopic binary UZ Tau E.  These results
come from the combination of our measurements of the mass ratio,
$M_2/M_1=0.28\pm$0.01, obtained using high resolution $H$-band spectroscopy,
with the total mass of the system, (1.31$\pm$0.08)(D/140pc) $M_{\odot}$,
derived from millimeter observations of the circumbinary disk \citep{sim00}.
The masses of the primary and secondary
are (1.016$\pm$0.065)(D/140pc) $M_{\odot}$
and (0.294$\pm$0.027)(D/140pc) $M_{\odot}$, respectively.  Using the
orbital parameters determined from our six epochs of observation, we
find that the inclination of the binary orbit, 59.8$\pm$4.4 degrees, is
consistent with that determined for the circumbinary disk from
the millimeter observations, indicating
that the disk and binary orbits are probably coplanar.

\end{abstract}


\section{Introduction}

This letter reports on the most recent progress in our efforts to
convert pre$-$main-sequence (PMS) single-lined
spectroscopic binaries (SB1s) to double-lined
systems (SB2s) in order to (1) provide dynamical mass
data for the calibration of young star evolutionary models, and
(2) to help determine the unbiased, low-mass PMS mass ratio
distribution of SB2s \citep{pra02}.  
Dynamical mass ratios have almost exclusively been measured 
in visible light and show a distribution weighted strongly
towards unity.  Because this is probably a selection
effect, given that the detection of a spectroscopic secondary is far
easier for flux, and therefore mass, ratios close to unity
\citep{maz02}, the
underlying astrophysical distribution of the mass ratio values
is unknown.  In \citet{pra02} we applied infrared (IR)
spectroscopy to the problem of identifying the secondary stars in
PMS SB1s because these low-mass companions are red
and thus are more readily detectable in IR light.
This project is an endeavor to convert all known, PMS SB1s
to SB2s.  \citet{pra02} show that this approach yields the
smallest mass-ratios ever derived for PMS SB systems.  To date,
all of the IR-converted SB2s display values of
$q=M_2/M_1<0.6$ \citep{ste01, pra02}.

We have applied this IR approach to UZ Tau E,
identified as a 19.1 day period SB1 by \citet{mat96}.  It
forms a hierarchical quadruple with UZ Tau W, itself 
a 0$\farcs$34 binary, $\sim$4$''$ to the west \citep{sim95}.
Together with GW Ori, DQ Tau \citep{mat91, mat97}, and V4046 Sgr
\citep{qua00}, UZ Tau E is one of only 4 classical T Tauri star SBs known.
Emission line and color diagnostics indicate that accretion is
occurring on to the stars in UZ Tau E \citep{ken95} despite
the expected formation of a gap in its circumbinary disk
\citep{art94}.  The presence
of the circumbinary disk around the UZ Tau E SB enabled
\citet{sim00} to measure the distance dependent, total binary
mass by mapping the Keplerian rotation of the $^{12}$CO gas in
the disk.  Thus, with the determination of the mass ratio, we
are able to calculate the component stellar masses and
orbital inclination.
In \S 2 we briefly describe our observations and
data reduction.  The analysis
and results appear in \S 3. Section 4 provides a brief discussion.

\section{Observations and Data Reduction}

$H$-band observations were made with the Keck II
near-IR spectrometer, NIRSPEC, a cross-dispersed, 
cryogenic echelle spectrometer employing a 1024 $\times$ 1024 ALADDIN 
InSb array detector \citep{mcl98, mcl00}.  The 
resolving power was $R = 24,000$ and $R = 30,000$ for the non-adaptive
optics (non-AO) and AO modes, respectively.  For AO mode observations,
dispersion solutions were derived using arc lamp lines. Otherwise
we used night sky OH lines \citep{rou00}.
Integration times for a single exposure were usually
300 s.  Further details about the observations are
provided in \citet{pra02}.

The data were reduced using REDSPEC, software designed at UCLA
for the analysis of NIRSPEC
data\footnote{See: 
http://www2.keck.hawaii.edu/inst/nirspec/redspec/index.html}.
For the analysis described here we used only order 49.
The central wavelength of this order is $\sim$1.55 $\mu$m; order 49
is almost completely free from terrestrial absorption lines.  
Table 1 provides a log of the observations; column (1) gives the UT date,
column (2) the observing mode (i.e. non-AO or AO), and column (3)
the modified Julian day of the observation.

\section{Analysis and Results}

Figure 1 shows the spectra from the six epochs of observations.
These were analyzed as described in \citet{maz02} and \citet{pra02}
using the two-dimensional cross-correlation
program TODCOR \citep{zuc94} and our library of stellar
templates.  For every epoch
of observation, the secondary star spectrum was detected; velocities
for the primary and secondary stars appear in columns (4) and (5)
of Table 1.  Uncertainties in the velocities reflect the addition
in quadrature of the internal uncertainty in the TODCOR analysis
and the $\pm$1 km s$^{-1}$ uncertainty between
our velocity reference frame, as estimated from the template star
radial velocities, and that of others \citep{pra02}.

The stellar templates that produced the maximum
correlation in the TODCOR analysis were either GL 763 or GL 752A for
the primary, and GL 213 or GL 402, rotationally broadened
to 25$-$30 km s$^{-1}$, for the secondary.  The Centre de Donn\'ees
astronomiques de Strasbourg (SIMBAD) lists
GL 763 and GL 752A with spectral types of M0 and M3, respectively,
however, GL 752A is probably misclassified. 
Figure 1 of \citet{pra02} shows the spectra in our template library;
at 1.55 $\mu$m, the spectrum of GL 752A appears to be earlier than M3
and is better matched to an M0 or M1.5 star.  We therefore regard the
best fitting primary templates as consistent with the M1
visible light spectral
classification given by \citet{ken95} for the entire UZ Tau E system.
The spectral types of the best fitting secondary templates are both M4.

The average 1.55 $\mu$m flux ratio is $H_2/H_1=0.47$.
As discussed in \citet{pra02}, the cross-correlation analysis yields
component velocities that are very reliable because they are based on 
many spectral lines.  However, estimates of spectral type and flux ratio
must be regarded as only representative because of the mismatch in
surface gravity and metallicity between the main-sequence star
templates and the PMS targets.

The SB1 results reported by \citet{mat96} give
the orbital period, $P=19.1$ days, the eccentricity, $e=0.28$, the
projected primary semi-major axis, $a_1sini=0.03$ AU, and the primary
semi-amplitude, $K_1=17$ km s$^{-1}$.
We derived the orbital elements of the UZ Tau E
SB2 by a least-squares minimization.  The phases which
appear in column (6) of Table 1 were calculated using our
value for the period derived from this procedure,
$P=19.048\pm$0.011 days.  Figure 2 shows the orbital fit to the observed
velocities and Table 2 lists our orbital elements in standard
notation following \citet{hei78}.  We derive
a mass ratio of $q=M_2/M_1=0.289\pm0.025$.  Our results are
in excellent agreement with the parameters in common
measured by \citet{mat96}.

By measuring the Keplerian rotation of the UZ Tau E circumbinary disk,
\citet{sim00} determined $M_{total}=(1.31\pm0.08)(D/140 pc) M_{\odot}$
for the binary.  This scales with
distance because it depends on the radial scale of the disk.
Combining $M_{total}$ with $q$,
derived here, we obtain the orbital inclination
($sini=0.864\pm0.039)(140$/D pc)$^{1/3}$, or $i=59.8\pm$4.4
degrees for $D=140$ pc \citep{ken94}, and the component masses,
$M_1=1.016\pm0.065 M_{\odot}$ and $M_2=0.294\pm0.027 M_{\odot}$.

\section{Discussion}

The inclination of the UZ Tau E circumbinary disk measured by its
$^{12}$CO J=2$-$1 rotation is, for $D=140$ pc, 56 $\pm$2 degrees;
the apparent projected inclination
of the disk in 1.3 mm continuum emission is 54 $\pm$3 degrees
\citep{sim00}.  This is consistent, within the uncertainties, with
our derived value, $i=59.8\pm$4.4 degrees ($D=140$ pc), for the orbit
of the spectroscopic binary.  The
range of values is $i=56-65$ degrees for $D=160-120$ pc.
This consistency between the orbital and circumbinary disk
inclinations indicates that the stellar orbit and the disk are
probably coplanar.

Figure 3 shows the components of UZ Tau E plotted on the H-R
diagram.  The M1 primary and M4 secondary were assigned 
effective temperatures of 3700$\pm$150 K and 3300$\pm$150 K, respectively,
from the conversion presented in Figure 5 of \citet{luh00}.
Using $H_{total}=8.46$ mag for the SB2, corrected
for $A_V=1.49$ mag \citep{ken95} and
apportioned according to the flux ratio, $H_2/H_1=0.47$ (\S 3),
we obtained the component $H$-band magnitudes.  
Applying the appropriate bolometric correction,
2.31 mag for the M1 and 2.44 mag for the M4 \citep{har94, tok00},
then enabled the calculation of the component luminosities,
$L_1=0.63^{+0.19}_{-0.17}$ $L_{\odot}$
and $L_2=0.28^{+0.09}_{-0.07}$ $L_{\odot}$.  The large uncertainties in
$L_1$ and $L_2$ are domianted by the $\pm$20 pc uncertainty in
the location of UZ Tau E along the line of sight to the Taurus SFR.
The luminosity ratio, $L_2/L_1=0.44\pm0.18$, is
approximately equal to the $H$-band flux ratio \citep{pra02}.

The PMS evolutionary tracks of \citet{bar98} and \citet{pal99}
are shown in Figure 3.  Both sets of tracks indicate that the system is
relatively young, $\sim$1$\times$10$^6$ years.  For both sets of tracks, the
secondary star lies within 1 $\sigma$ of the mass track appropriate
to its derived dynamical mass; however, the primary star appears on mass
tracks with a value 3$-$4 $\sigma$ smaller than its dynamically
derived mass.  The model-based mass ratios, $q=0.52\pm0.23$ for
the tracks of \citet{bar98} and $q=0.38\pm0.23$ for those of \citet{pal99},
are consistent with the dynamical mass ratio, $q=0.29\pm0.03$ to within
1 $\sigma$, but this is a result of the propagation of the uncertainties
in the track-derived masses, which are as high as $\sim$50 \%.

It is unlikely that the source of this discrepancy
lies in the simple application of the $H$-band flux
ratio to apportion the component luminosities, even
though this ratio is uncertain (\S 3), because the mass of
an M1 star is relatively insensitive to
luminosity on the tracks of both \citet{bar98} and \citet{pal99}.
We can rule out contamination by a third component in the UZ Tau E
system on the basis of our cross-correlation analysis as well as
mm-wave and near-IR imaging of the system (e.g., Dutrey et al. 1996).

Several origins for this discrepancy are possible.
(1) UZ Tau E may be
on the near side of the Taurus star forming region at a distance
of $\sim$120 pc. The total mass of the SB2 system as derived by
\citet{sim00} is a function of distance.
(2) The mass of the spectroscopic system
measured by \citet{sim00} may be an overestimate if a dense ring
of circumbinary material with a radius of a few AU is present in the disk.
(3) The rotation of the circumbinary disk may be non-Keplerian.
(4) The main-sequence templates may be poorly matched to the
PMS objects and hence yield incorrect spectral types.
(5) Given the complexities of the calculations of PMS evolution at
very young ages, some uncertainties may also be expected in the
theoretical tracks.  

To test the plausibility of the first two possibilities listed above
we combine a 20 pc (14 \%) underestimate in
the distance to UZ Tau E and a 5 \% overestimate in the total stellar
mass measured at millimeter wavelengths, resulting from the
presence of an undetected dense ring of material within the
UZ Tau E circumbinary disk, similar to the structure in
the circumbinary disk of GG Tau \citep{gui99}.
We now derive $M_{total}=1.07\pm0.07$ $M_{\odot}$,
yielding $M_1=0.83\pm0.05$ $M_{\odot}$
and $M_2=0.24\pm0.02$ $M_{\odot}$.  The revised dynamical value of $M_1$
still departs by 2$-$3 $\sigma$ from the
location of the star on the H-R diagram.  The difference in the
dynamical and model values for $M_2$ is still $<$1 $\sigma$.

The low-surface gravity expected in a young, PMS star may cause
its spectrum to appear to be later than the object's mass would
indicate \citep{whi99, luh00}.  However, precisely to
account for such an effect, we have used the 
spectral type to effective temperature conversion of the \citet{luh00}
to place the SB2 components on the H-R diagram.  In addition,
agreement between the tracks and the dynamically measured
secondary star mass is inconsistent with this interpretation.
On the Baraffe tracks, the UZ Tau E secondary appears to be slightly {\it more}
massive than expected from the dynamical data.  On the tracks of
Palla \& Stahler, it is slightly less massive than expected, but for both
sets of tracks, the discrepancy in the position of the secondary is $<$1
$\sigma$ (Figure 3).  To investigate
this phenomenon further, observations and analyses of the type we report
here are required to connect stars of well-determined mass with the
appropriate spectral types and effective temperatures.

DQ Tau is also a short period (16 day) classical T Tauri SB2
\citep{mat97}.  It is remarkable in that accretion onto the central
stars from a circumbinary disk is regulated by the
eccentric orbit of the stars.  Most photometric studies of
UZ Tau E have not isolated the system from UZ Tau W; it is unclear
if photometric
variability synchronized with the period of the SB2 is present.
If the discrepancy between our
dynamical mass measurement and the masses obtained from the
H-R diagram is attributable to unusual accretion processes,
it is difficult to understand how such would cause a star to
appear {\it less} massive, i.e. appear to have
a later spectral type, than the dynamical measurement implies.

\citet{qua00} studied the components in another classical T Tauri
SB2, V4046 Sgr, a 2.4 d period system, and found a discrepancy
between the mass ratio determined dynamically, $q=0.94$, and from an
evolutionary model, $q=0.80$.  However, deriving uncertainties for their
observed and theoretical mass ratio, from their Table 1 and
Figure 3, respectively, we find that their numbers are consistent
to within 1 $\sigma$ and therefore are not discrepant.  The total mass
of the V4046 Sgr SB2 has not been measured dynamically so the
individual stellar masses are not known.

\citet{qua00} suggest that the unusual behavior in V4046 Sgr might be
attributable to the distorted structure of stars in such a short period
binary.  However, in UZ Tau E, with an orbital period of $\sim$19 d,
the tidal effects will be much smaller.  With an orbital semimajor axis
of $\sim$33 $R_{\odot}$, there is only a weak interaction between the
2$-$3 $R_{\odot}$ radius components.
Unfortunately, it is not possible at the present time to deconvolve
the effects of metallicity, rotation, and surface gravity 
simultaneously for blended components of an SB2.  A combination of
some of the factors discussed above may ultimately be responsible
for the inconsistencies discussed in this letter.  
Clarification of the questions raised here may
require very high spatial resolution orbital mapping with future
facilities such as the {\it Space Interferometry Mission}.

\section{Acknowledgements}

Almost every single Keck observing assistant helped with some
epoch of these observations; we are grateful for their expertise. 
We thank the staff and support scientists for their logistical
and technical assistance, in particular, Barbara Schaefer,
Randy Cambell, and David Le Mignant for their help with the
2001, August 31 service observations.  We thank Anne Dutrey and
Stephane Guilloteau for helpful discussions, and an anonymous
referee for comments which improved this paper.
This research was supported in part
by NSF Grants AST 98-19694 and AST 02-05427 (to M. S.).
Data presented herein were obtained at the W. M. Keck 
Observatory, which is operated as a scientific partnership
between the California 
Institute of Technology, the University of California, and NASA.
The Observatory was made possible by the generous financial support of the 
W. M. Keck Foundation.  The authors wish to extend special thanks to those
of Hawaiian ancestry on whose sacred mountain we are
privileged to be guests.  This research has made use of the SIMBAD
database, operated at CDS, Strasbourg, France.

\clearpage

\pagestyle{empty}

\begin{deluxetable}{lccrrl}
\tablewidth{0pt}
\tablecaption{Summary of Observations and Analysis\label{tbl-1}}
\tablehead{
\colhead{UT Date of} & \colhead{Mode of} & \colhead{Modified Julian Day} &
\colhead{$v_{primary}$}  & \colhead{$v_{secondary}$  }  & \colhead{$~$} \\
\colhead{Observation} & \colhead{Observations} & \colhead{(2,450,000$+$)} &
\colhead{(km s$^{-1}$)} & \colhead{(km s$^{-1}$)} & \colhead{Phase}}
\startdata
2000 Nov 11 & non-AO & 1859.6  & 5.8$\pm$1.04 & 44.2$\pm$1.49  & 0.776 \\
2001 Jan 5 & AO & 1914.2  & 15.0$\pm$1.12 & 11.7$\pm$1.64 &  0.643 \\
2001 Aug 4 & non-AO & 2125.6  & 8.5$\pm$1.04 & 39.8$\pm$1.49 &  0.745 \\
2001 Aug 31 & AO & 2152.6  & 21.8$\pm$1.04 & -11.0$\pm$1.72 &  0.163 \\
2001 Oct 10 & non-AO & 2192.6  & 27.2$\pm$1.08 & -23.7$\pm$2.33 &  0.264 \\
2002 Feb 6 & non-AO & 2311.2  & 27.4$\pm$1.12 & -31.5$\pm$1.89 &  0.492 \\
\enddata

\end{deluxetable}

\clearpage

\pagestyle{empty}

\begin{deluxetable}{lc}
\tablewidth{0pt}
\tablecaption{Orbital Elements and Derived Properties
of UZ Tau E \label{tbl-2}}
\tablehead{}
\startdata
$P = 19.048 \pm 0.011$ days\\
$\gamma = 14.8 \pm 0.6$ km s$^{-1}$\\
$K_1 = 17.4 \pm 1.4$ km s$^{-1}$\\
$K_2 = 60.2 \pm 3.0$ km s$^{-1}$\\
$e = 0.237 \pm 0.030$ \\
$\omega = 220.5 \pm 7.6$ degrees\\

\medskip
$T = 2,452,092.45 \pm 0.44$ MJD\\
$M_1 sin^3 i = 0.655 \pm 0.098 M_{\odot}$\\
$M_2 sin^3 i = 0.190 \pm 0.031 M_{\odot}$\\
$q = M_2/M_1 = 0.289 \pm 0.025$ \\
$a_1 sin i = (4.42 \pm 0.35) \times 10^6$ km\\

\medskip
$a_2 sin i = (15.31 \pm 0.71) \times10^6$ km\\
$M\tablenotemark{a}_{total} = 1.31 \pm 0.08 M_{\odot}$ for D$=$140 pc\\
$M_1 = 1.016 \pm 0.065 M_{\odot}$ for D$=$140 pc\\
$M_2 = 0.294 \pm 0.027 M_{\odot}$ for D$=$140 pc\\
$i = 59.8 \pm 4.4$ degrees for D$=$140 pc\\
\enddata

\tablenotetext{a}{Data from \citet{sim00}}
\end{deluxetable}

\clearpage

\figcaption[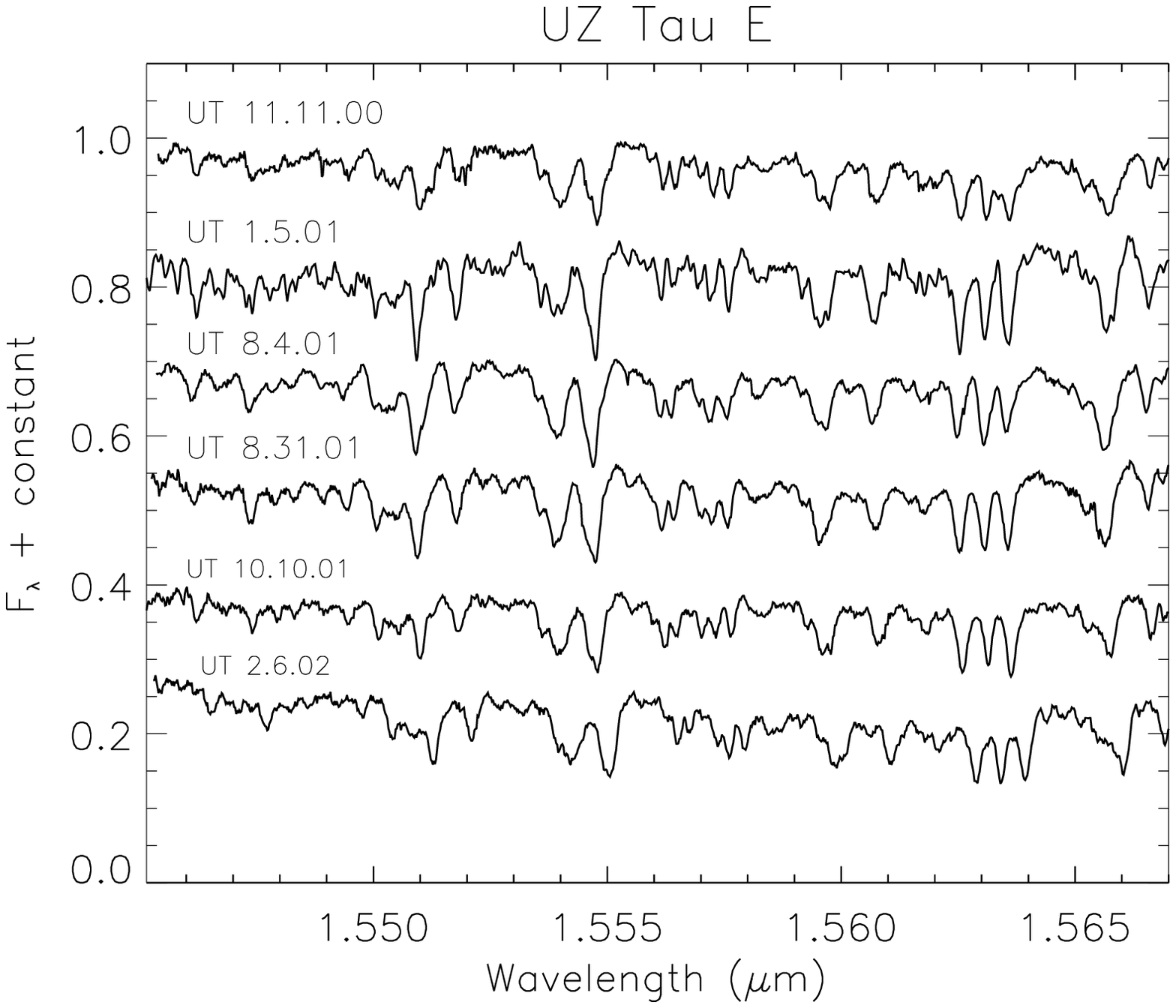]{Six epochs of order 49 NIRSPEC spectra of
the UZ Tau E spectroscopic binary.
No heliocentric or radial velocity corrections have been applied. 
The spectral continuum has been flattened.}

\figcaption[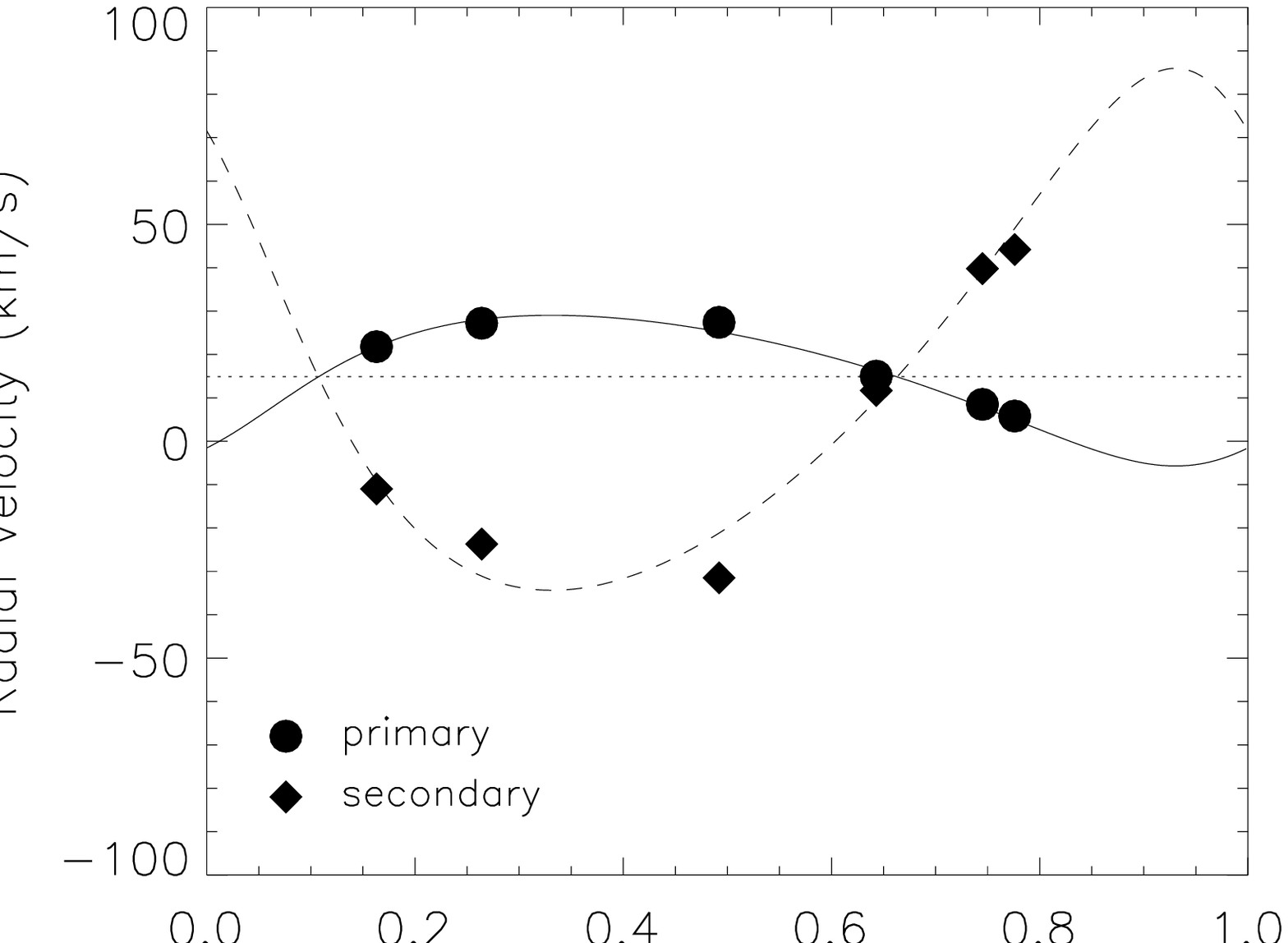]{Radial velocity as a function of phase
for UZ Tau E.  The circles represent primary star data and
the diamonds secondary star data.  The best fit to the
data is shown as a solid line for the primary star and a
dashed line for the secondary.  A dotted, horizontal line 
indicates the center of mass velocity of the system.  The
uncertainties in the velocities (Table 1) are smaller than
the plotting symbols.}

\figcaption[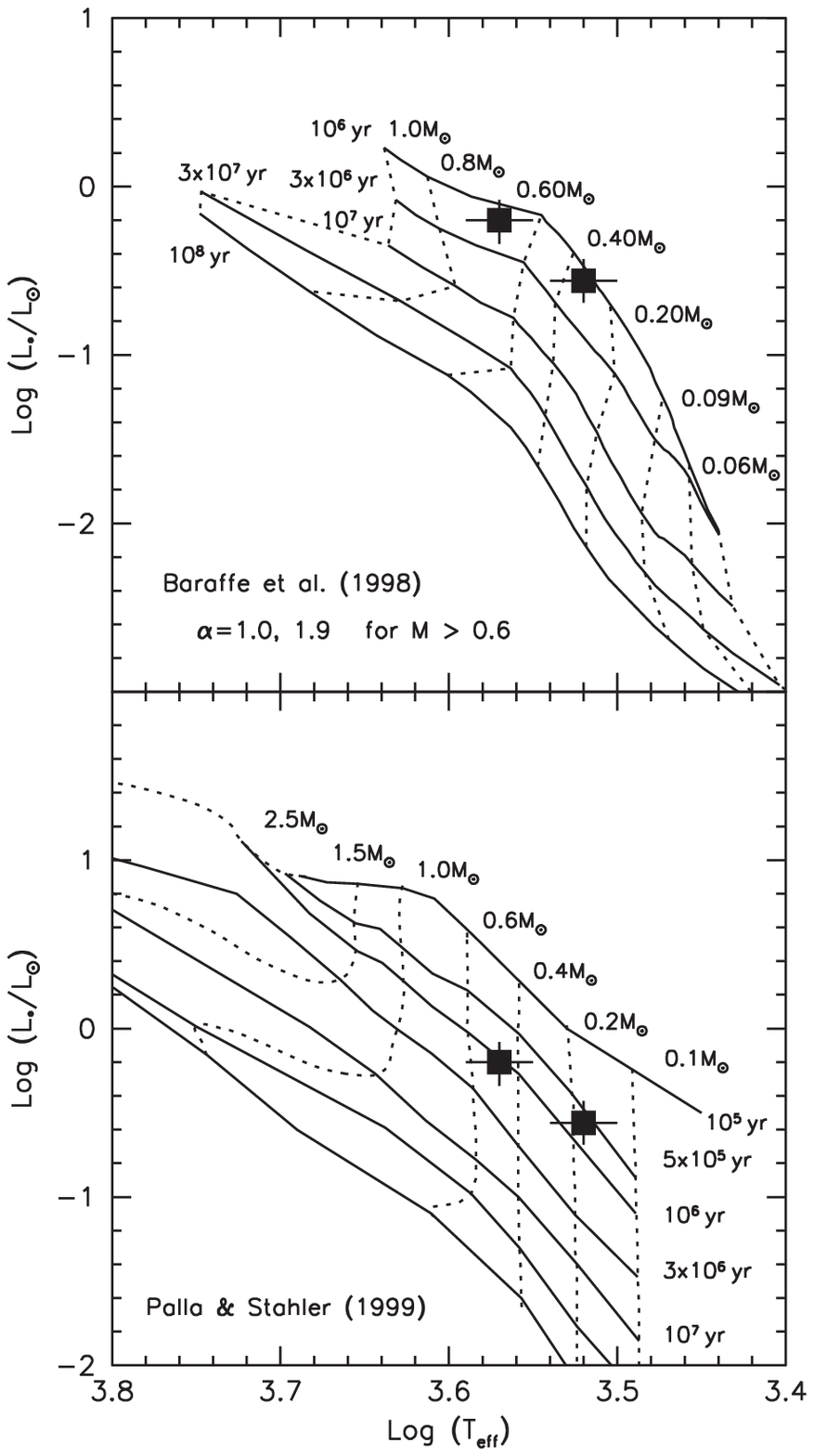]{The H-R diagram showing the components of
the UZ Tau E system on the tracks of \citet{bar98}
and \citet{pal99}.  The spectral type to temperature conversion
for the component M1 and M4 stars was made using the
scale defined by \citet{luh00}.  Uncertainties in $T_{eff}$
correspond to one spectral subclass.  The derivation of the
luminosities is described in the text.  In the upper panel,
the mixing length parameter, $\alpha$, is 1.0 for $M<0.6 M_{\odot}$
and 1.9 for $M>0.6 M_{\odot}$.}

\clearpage

\begin{figure}
\figurenum{1}
\plotone{f1.eps}
\end{figure}

\begin{figure}
\figurenum{2}
\plotone{f2.eps}
\end{figure}

\begin{figure}
\figurenum{3}
\plotone{f3.eps}
\end{figure}

\end{document}